\documentstyle[preprint,epsfig,aps,prabib]{revtex}

\begin{document}

\draft

\title{Photodisintegration of the triton \\
with realistic potentials }

\author{W. Schadow and W. Sandhas}
 \address{Physikalisches
Institut der Universit\"at Bonn\\ Endenicher Allee 11-13\\ 53115
 Bonn, Germany}

\maketitle

\begin{abstract} The process $\gamma + t \to n + d$ is
treated by means of three-body integral equations employing
in their kernel the {\sl W}-Matrix
representation of the subsystem amplitudes. As
compared to the plane wave (Born) approximation the full solution
 of the integral equations, which takes into account the final
  state interaction, shows at low energies a 24\% enhancement.
The calculations are based on the semirealistic Malfliet-Tjon
and the realistic Paris and Bonn {\sl B} potentials. For comparison with
earlier calculations we also present results for the Yamaguchi
 potential. In the low-energy region a remarkable potential
  dependence is observed, which vanishes at higher energies.
\end{abstract}


\newpage

\narrowtext

\section{Introduction}

The photodisintegration of $\,{^3}$H and ${^3}$He and the inverse
 reaction, the radiative capture of protons or neutrons by deuterons, have
been intensively investigated in the past. Due to the fact that the
 corresponding matrix elements contain both the three-body bound and
continuum states, these reactions are expected to be
 a good testing ground for the underlying two-body potential.
In  early calculations by Barbour and Phillips \cite{Barb70} it has
 been shown that an exact
treatment of the continuum states results in a considerable enhancement
 of the cross section in the peak region.
In that work and in the following by Gibson and Lehman \cite{Gibs75} only
 simple $s$-wave interactions of Yamaguchi type have been used,  while
 $d$-components $(j \leq 1^+)$  were incorporated in \cite{Fons91a}.
The role  of  $p$-wave contributions in the two-body input  with respect
 to the three-body cross section was
investigated by Fonseca  and Lehman \cite{Fons92b}, using again
 Yamaguchi terms in the  interaction.
More recently, calculations based on separable representations of
the Bonn {\sl A}  and Paris  potentials,
including their higher partial wave contributions, were performed for
 polarization observables at some specific energies
\cite{Fons93a,Ishi92a,Fons95a,Schmi96a}.

 In the present work we  calculate the differential cross sections at $90^o$
for the Yamaguchi, Malfliet-Tjon, Paris, and Bonn {\sl B} potentials
 over an energy region from threshold up
to $40$ MeV. The potential dependence, the effect of higher partial waves,
 and the role of meson exchange currents, taken into account via Siegert's
theorem, are investigated.

 Technically the calculations are based
on the Faddeev-type Alt-Grassberger-Sandhas (AGS) formalism \cite{Alt67}
adjusted to photonuclear processes \cite{Sand86}, as done already
in \cite{Gibs75}.
The separable representation of the subsystem {\sl T} matrices, relevant in
this context , is chosen according to the {\sl W}-Matrix approach
 \cite{Bart86}.
This approximation  combines
high accuracy with considerable simplicity \cite{Janu93}.

A more reliable, still fairly simple representation
of the two-body input is provided by the Ernst-Shakin-Thaler method
 \cite{Erns73}.
Calculations employing the corresponding  PEST, BEST, and NEST
 potentials \cite{Haid84,Haid86b,Haidpriv}
will be presented in a subsequent publication.

\section{Formalism}

The  Alt-Grassberger-Sandhas (AGS) equations are well known to go over
 into effective two-body
Lippmann-Schwinger equations \cite{Alt67} when representing the input
two-body {\sl T}-operators in separable form. The neutron-deuteron
off-shell
scattering amplitude ${\cal T} (\vec q, \vec q\,'')$, thus, is
 determined by

\begin{equation}
\label{scampl}
 {\cal T} (\vec q, \vec q\,'') = {\cal V} (\vec q, \vec q\, '')
 + \int \! d^{\,3} q' \: {\cal V} (\vec q, \vec q\,')\:{\cal G}_0
 (\vec q\,') \: {\cal T} (\vec q\,',\vec q\,'') .
\end{equation}

\noindent
Applying the same technique to the photodisintegration of the triton
into neutron and deuteron, i.e. to the process $ \gamma + t \rightarrow
 n + d$, an integral equation of rather similar structure is obtained
 \cite{Barb70,Gibs75,Sand86},

\begin{equation}
\label{pampl}
{\cal M} (\vec q\,) = {\cal B} (\vec q\,) + \int \! d^{\,3} q' \:
{\cal V}
(\vec q,\vec q\,') \:{\cal G}_0 (\vec q\,') \: {\cal M}(\vec q\,') \;,
\end{equation}

\noindent
 where ${\cal M}(\vec q)$ represents an off-shell extension
   of the full photodisintegration amplitude

\begin{equation}
\label{mampl}
 M (\vec q\,) = \;  ^{(-)} \langle\vec q;  \psi_{d}|
 H_{\text {em}}|\Psi_t\rangle \;.
\end{equation}

In both these equations the kernel is given by the effective
neutron-deuteron potential ${\cal V}$ and the corresponding
effective free Green function ${\cal G}_0$ defined in Ref.
\cite{Alt67}. However, in Eq. (\ref{pampl})
the inhomogeneity of Eq. (\ref{scampl}) is replaced by an
 off-shell extension ${\cal B} (\vec q\,)$ of the plane-wave (Born)
  amplitude

\begin{equation}
 B (\vec q \,) = \: \langle\vec q\,|  \langle \psi_{d}| H_{\text {em}}|\Psi_t
 \rangle \;.
\end{equation}

\noindent
Here  $ \left| \Psi_t \right\rangle $ and  $\left| \psi_d \right\rangle $
 denote the triton and deuteron states,  $\left| \vec q\; \right\rangle $ the
momentum state of the neutron relative to the center of mass of the
deuteron, and $H_{\text {em}}$  the electromagnetic operator.
Thus, by this replacement any working program for
$n$--$d$ scattering can immediately be applied to the above
photoprocess.

\bigskip

The results presented in this paper are, in fact, obtained
by extending recent $n$--$d$ calculations \cite{Janu93,Bart87}
in this manner. As in these references, the separable part of
the {\sl W}-matrix
representation \cite{Bart86} of the two-body {\sl T} matrices is
employed,

\begin{equation}
  T^\eta_{ll'} (p,p', E + i0) =
 \sum_{\hat l \hat l'} p^l W^\eta_{l \hat l} (p, k; E)
\Delta^\eta_{\hat l \hat l'} (E + i0) W^\eta_{l' \hat l'} (p', k; E)
 p'^{l'} \;.
\end{equation}

\noindent
Here $l$ and $l'$ are the orbital angular momenta, and $\eta = (s,j;i)$
stands for the spin, the angular momentum j [ with the coupling
sequence $(l,s)j$] and the isospin $i$ of the two-body subsystem.
With the form factors $W_{\hat l l}^\eta$ and the propagators
$\Delta^{\eta}_{\hat l \hat l'}$ of this representation Eq.
(\ref{pampl}) reads, after partial wave decomposition and
 antisymmetrization,

\widetext
\begin{equation}
 \label{phampl} ^{\Gamma}{\cal M}^{b }_{\hat l}
(q,E) = \,^{\Gamma}{\cal B}^{b}_{\hat l}(q,E) + \sum_{b' \hat l',
\hat l''}  \int \limits_0^\infty dq' q'^2 \,\,^{\Gamma}{\cal V}^
{b b'}_{\hat l \hat l'} (q',E)  \Delta^{\eta'}_{\hat l \hat l'}
(E - \textstyle \frac{3}{4}\, q'^2) \,^{\Gamma}{\cal M}^{b'}_{\hat l''}(q',E)
\end{equation}

\noindent
and its inhomogeneity is given by

\begin{equation}
\label{bampl}
  ^{\Gamma}{\cal B}_{\hat l}^{b } (q, E) =
  {\sqrt 3} \sum_{ l} \int \limits_0^\infty dp \, p^{l + 2}
  \langle \, p q b l \Gamma I  \,|\, W_{l \hat l}^\eta
(k,p, E - \textstyle \frac{3}{4}\, q^2) \, G_0 (E + i\,0) H_{\text {em}}
 \,|\,\Psi_t \, \rangle \;.
\end{equation}

\narrowtext

\noindent
Here, $ G_0$  and $\left\langle  p q b l \Gamma I \right|$
 represent the
 free three-body
 Green function and the partial wave projection of the three-body plane
 wave state $\left\langle \vec q\,\right | \! \left\langle \vec p\, \right |$.
 The detailed
structure of the effective potential $^{\Gamma }{\cal V}^{b b'}
_{\hat l \hat l'}$ can be found
in Ref. \cite{Janu93}.
 The label $b$  denotes the set $(\eta K L)$ of quantum
numbers, where $K$ and $L$ are the
channel spin of the three nucleons [ with the coupling sequence $(j,
\frac{1}{2}) K$] and the relative angular momentum between the two-body
subsystem and the third particle, respectively.  $\Gamma$ is the total
 angular momentum following from the coupling sequence $(K,L) \Gamma$,
and $I$ is the total isospin.

\bigskip

The Born amplitude contains the triton bound state $\left| \Psi_t \right\rangle $,
 which may be calculated by any of the various bound-state methods.
Consistently with the present approach we employ for this purpose
the partial wave projected homogeneous version of Eq. (\ref{scampl}),

\widetext
\begin{equation}
  F^{b}_{\hat l} (q) = \sum_{b' \hat l' l} \int \limits^\infty_0 dq' \,
   q'^2 \,
 {\cal V}^{b b'}_{\hat l \hat l'} (q, q', E_T) \, \Delta^\eta_{l \hat l}
 (E_T - \textstyle \frac{3}{4}\, q'^2)  \, F^{b'}_{\hat l'} (q').
\label{form}
\end{equation}

\narrowtext
\noindent
Its solutions, the three-body ``form-factors'' $F^{b}_{l} (q)$, are
related to $\left| \Psi_t \right\rangle $ according to \cite{Alt67} by

\begin{equation}
\left| \Psi_t \right\rangle  =  \sum_{\gamma b l \hat l \hat l'}
\int \! \! \int dq \, dp \, q^2 \, p^2 \,G_0(E_T)
\, \left| (\gamma) q p b l \Gamma I \right\rangle
 W_{l \hat l} (p,k,E_T - \textstyle \frac{3}{4}\, q^2) \,
\Delta^\eta_{\hat l \hat l'} (E_T - \textstyle \frac{3}{4}\, q^2) \, F^b_{\hat l'} (q).
\end{equation}

\noindent
Here, the summation runs over all two-fragment partitions $\gamma$,
the variables and quantum numbers being understood in the corresponding
set of Jacobi coordinates.

\bigskip

The electromagnetic operator entering Eq. (\ref{bampl})
is, at the low energies considered, essentially a dipole operator.
Ignoring meson-exchange currents it is given by

\begin{equation}
\label{Sieg1}
 H'_{\text {em}} = \sqrt{\frac{4 \pi}{3}} \,
\sum_{i = 1}^3 e_i  \, p_i \, Y_{1 \lambda}(\vartheta, \varphi) \;.
\end{equation}

\noindent
With exchange currents it takes,
according to Siegert's theorem \cite{Sieg37}, the form
\cite{Gian85}

\begin{equation}
\label{Sieg2}
 H_{\text {em}} = -i \,\sqrt{\frac{4 \pi}{3}} \, (E_f - E_t )\,
\sum_{i = 1}^3 e_i  \, r_i \, Y_{1 \lambda}(\vartheta, \varphi) \;.
\end{equation}

\noindent
Here $E_f$ and $E_t$ denote the final and the triton energies,
$r_i$ the nucleon center-of-mass coordinates, $p_i$ the corresponding momenta
, $e_i$ the electric charges, and $\lambda$ the polarization of the photon.
\bigskip

The on-shell restricted  solutions  $^{\Gamma}{ M}^{b }_{\hat l}$ of the
 integral Eq. (\ref{phampl}) yield the photodisintegration amplitude via the
partial-wave summation

\widetext
\begin{eqnarray}
 \lefteqn{ ^{(-)}
\langle \, \vec q s m_s; \psi_d j m_j \, | \, H_{\text {em}} \,|\,
\Psi_t \, \Gamma'  M_{\Gamma'} \,\rangle }
  \nonumber \\ & =  & \sum_{\Gamma M_{\Gamma} b \, \hat l M_K M_L}
  \langle \, j m_j s m_s \,|\, K M_K \, \rangle
  \langle \, K M_K L M_L \,|\, \Gamma M_{\Gamma} \, \rangle \;
   Y_{L M_L} ( \hat q) \,\,  {^{\Gamma}}\! {M}^{b} _{\hat l}\;
   (q, E_d + \textstyle \frac{3}{4}\, q^2) \;,
\end{eqnarray}

\narrowtext
\noindent
where now, in contrast to Eq. (\ref{mampl}), the spin and angular
momentum quantum numbers of the neutron ($s m_s$), deuteron ($j m_j$),
 triton ($\Gamma' M_{\Gamma '}$), and
the polarization ($\lambda$) of the photon are explicitly given.
Denoting this amplitude  by $M(\vec q \,)_{m_s m_j M_{\Gamma '}
\lambda}$ ,
the cross section is obtained in the standard way by

\begin{eqnarray}
 \frac{d\sigma}{d\Omega} = \frac {2 \pi^2}{3}
 \frac{q}{E_\gamma \, c} \sum_{m_s m_j} \sum_{ M_{\Gamma'} \lambda}
 \left |\, M(\vec q \,)_{m_s m_j M_{\Gamma'} \lambda}  \right |^2 \;.
\end{eqnarray}

\noindent
Here we have averaged over
the initial states and summed over the final states.

\section{Results}

As a first test of our numerical program we  performed
calculations for the Yamaguchi \cite{Yama54} potential in order to compare
 them with the corresponding results  by Gibson and Lehman  \cite{Gibs75}.
 Unfortunately, employing the same sets
of parameters as in this reference, we read off from
the analytically given relations somewhat different values for the
two-body scattering length, effective range, and
deuteron binding energy (Table I).
The triton binding energies,
obtained by solving the homogeneous three-body equations, differ
 in one case,  the symmetric Tabakin potential, where we found
$-9.72$MeV instead of $-9.33$MeV (Table III).  These
 discrepancies
mean that our photodisintegration results cannot be expected
to fully agree with the ones of Ref. \cite{Gibs75}.

 Figure 1 compares our cross sections for  triton-photodisintegration
  (solid line)
with the calculations by Gibson and Lehman \cite{Gibs75} (dashed line).
The disagreement, in particular in the peak region, is
reduced when replacing in the Siegert operator  our triton
energy   by the Gibson-Lehman
value (short-dashed line). Figure 2
shows the same comparison, but now for the $^3$He photodisintegration.
Since there is no disagreement between the $^3$He binding energies, a
correspondingly
modified curve  does not exist. The remaining discrepancies,
therefore, have to be attributed to numerical uncertainties,
 which are not unexpected in view of  the level of accuracy
 reached in early calculations. Within these
limits we consider our results for the Yamaguchi potential consistent with
the ones of  Ref. \cite{Gibs75}.

The differential cross sections  obtained for the Paris \cite{Laco80},
Bonn {\sl B} \cite{Mach87}, Malfliet-Tjon (MT I-III) \cite{Malt70}, and
 Yamaguchi potentials are shown in Figure 3. Most remarkable
is the strong potential dependence in the peak region, which vanishes
at higher energies. A further observation, which should be relevant
in model calculations or when going over to higher particle
numbers \cite{Eller96}, is the proximity of the Paris and Malfliet-Tjon
results on the one hand, and less closely the Bonn {\sl B} and Yamaguchi potentials
on the other hand.

In Figure 4 we contrast, for the Paris potential, the solution
of the integral equation (solid line) with the corresponding plane-wave
(Born) approximation (dashed line). It is seen that the full  solution
is enhanced by about 24\% at the peak. A similar enhancement was
observed for simpler interactions already in Ref. \cite{Barb70}.
The upper curves are the ones based on the Siegert-operator
(\ref{Sieg2}), the lower ones correspond to the non-Siegert
 operator (\ref{Sieg1}).  There is a factor of two between the Siegert
and non-Siegert results,
 which demonstrates the relevance of meson-exchange currents.
For a detailed discussion  of  the same phenomenon
in case of deuteron and $^4$He photodisintegration
we refer to Refs. \cite{Aren91} and \cite{Eller96}, respectively.

Figures 5 and 6 compare our  cross sections for the
s- and d-wave projected Paris and Bonn {\sl B} potentials with the
 data  from Refs. \cite{Kosi66,Faul80,Skop81}.   Figures 7 and 8 show
the same with inclusion of the subsystem $p$ waves,
which leads to a reduction of the peak by 8--10\%.
Up to 25 MeV the best agreement with the data is achieved for
the Bonn {\sl B} potential.
In view of the experimental errors the relevance of this observation
is, of course,  somewhat questionable.
At higher energies the potential dependence vanishes. It is, however,
seen that
the incorporation of the $p$ waves is essential for the
remarkable agreement with most recent experimental data
\cite{Skop81}. Note that for the inverse reaction the relevance
of the $p$ wave contributions has been pointed out also
in \cite{Fons93a,Ishi92a,Fons95a,Schmi96a}.

Two fragment photodisintegration of the three-nucleon bound states, thus,
provides a  sensitive tool for testing the underlying two-nucleon potentials.
A repetition of the corresponding low-energy measurements with much higher
accuracy is strongly suggested by this observation.

\begin{acknowledgments}
This work  was supported by the
Deutsche Forschungsgemeinschaft under Grant No. Sa 327/23-1.
\end{acknowledgments}

\begin{table}[h]
\caption{Parameters for the Yamaguchi-Potential. The numbers in
parenthesis are taken from [2].}
\begin{tabular}{lcccccc}
              & & & & & &  Binding\\
              & & Strength & Inverse range & Scattering length &
Effective range & energy \\
\multicolumn{1}{c}{Interaction} & & $\lambda \;({\text {fm}}^{-3})$ & $ \beta
\; ({\text {fm}}^{-1})$ & $ a \;({\text {fm}}) $ & $ r_0 \;({\text {fm}})$ &
 (MeV)\\
\hline
 {\sl n-p} triplet & I  & 0.3815 & 1.406 &  5.433  &  1.761 & -2.203\\
                   &    &        &       & (5.423) & (1.761)& (-2.225) \\
                   & II & 0.220  & 1.15  &  5.806  &  2.088 &  -2.082\\
                   &    &         &      & (5.68)  & (2.09) & (-2.225)\\
 {\sl n-p} singlet & I  & 0.1445 & 1.153 & -23.196   &  2.732 & \\
                   &    &        &       & (-23.715) & (2.74) &\\
                   & II  & 0.148 & 1.15 & -42.217   &  2.680  &\\
                   &    &        &       & (-21.25) & (2.74) &\\
 {\sl p-p} singlet &   & 0.1534 & 1.223 & -7.853   &  2.794  &\\
                   &    &        &       & (-7.823) & (2.794) & \\
 {\sl n-n} singlet &   & 0.1323 & 1.130 & -16.851   &  2.841  &\\
                   &    &        &       & (-17.0) & (2.84) &\\
\hline
 {\sl n-p} triplet & III & -0.220 & 1.14525 & 5.666 & 2.081 & 2.226  \\
 {\sl n-n} singlet & III & -0.148 & 1.16225 &  -22.998 & 2.710 & \\
\end{tabular}
\end{table}

\begin{table}[h]
\caption{Parameters for the Malfliet-Tjon-Potential.}
\begin{tabular}{|@{\hspace{5mm}}c@{\hspace{5mm}}|
@{\hspace{6mm}}c@{\hspace{6mm}}|@{\hspace{6mm}}c@{\hspace{6mm}}|
@{\hspace{6mm}}c@{\hspace{6mm}}|@{\hspace{6mm}}c@{\hspace{6mm}}|}
 $\sigma$ & $\lambda_{\sigma}^A [{\text {fm}}^{-2}]$ & $\lambda_{\sigma}^R
[{\text {fm}}^{-2}]$ &   $r_{\sigma}^A [{\text {fm}}]$ & $r_{\sigma}^R
 [{\text {fm}}]$ \\
\hline
 $s$ &  -19.5719  &  &  &  \\
 $d$ &  -23.8775  & \raisebox{3ex}[-3ex]{109.605} &
                    \raisebox{3ex}[-3ex]{0.643087} &
                    \raisebox{3ex}[-3ex]{0.321543} \\
\end{tabular}
\end{table}

\begin{table}[h]
\caption{Triton Binding energies obtained with the Yamaguchi-Potentials.
The numbers in parenthesis are taken from [2].}
\begin{tabular}{ccc}
                 &                     & Binding energy \\
 Wave function   & Interaction set     & (MeV) \\
\hline
 Symmetric Tabakin      & Average of $n-p$    &  9.72  \\
                 & triplet and singlet & (9.33) \\
                 & sets II             &        \\
Tabakin          & $n-p$ triplet and   &  10.11 \\
                 & singlet sets II     &  (10.1) \\
Charge dependent & $n-p$ triplet I     &  10.34  \\
 $^3$H           & $n-p$ singlet I     &  (10.34) \\
                 & $n-n$ singlet       &         \\
Charge dependent & $n-p$ triplet       &   8.49  \\
                 & $\lambda = 0.3608 \,{\text {fm}}^{-3}$ & (8.49) \\
 $^3$H           & $n-p$ singlet I     &   \\
 (adjusted)      & $n-n$ singlet       &         \\
Charge dependent & $n-p$ triplet I     &  7.72  \\
                 & $\lambda = 0.3589 \,{\text {fm}}^{-3} $& (7.72) \\
 $^3$He           & $n-p$ singlet I     &   \\
 (adjusted)      & $p-p$ singlet       &         \\
\hline
 $^3$H           & $n-p$ singlet III   &  9.968 \\
 (present)       & $n-n$ singlet III   &         \\
\end{tabular}
\end{table}

\newpage

\begin{center} FIGURES \end{center}

\noindent
{FIG. 1.} Triton photodisintegration cross section for the
Yamaguchi potential compared with [2].

\vspace{1cm}

\noindent
{FIG. 2.} Same as Fig. 1 but for $^3$He.

\vspace{1cm}

\noindent
{FIG. 3.} Triton photodisintegration cross sections for the Yamaguchi,
Malfliet-Tjon, Paris and Bonn {\sl B} potentials.

\vspace{1cm}

\noindent
{FIG. 4.} Triton photodisintegration  cross sections for the Paris
potential ($j \leq 1^+$) with and without Siegert's  theorem.

\vspace{1cm}

\noindent
{FIG. 5.} Cross sections for the Paris potential ($j \leq 1^+$) compared
with the data  from Refs. [23-25].

\vspace{1cm}

\noindent
{FIG. 6.} Same as Fig. 5 for the Bonn {\sl B} potential.

\vspace{1cm}

\noindent
{FIG. 7.} Cross sections for the Paris potential with incorporation of
the $p$ waves   ($j \leq 1$) compared
with the data  from Refs. [23-25].

\vspace{1cm}

\noindent
{FIG. 8.} Same as Fig. 7 for the Bonn {\sl B} potential.

\newpage

\hbox{\psfig{figure=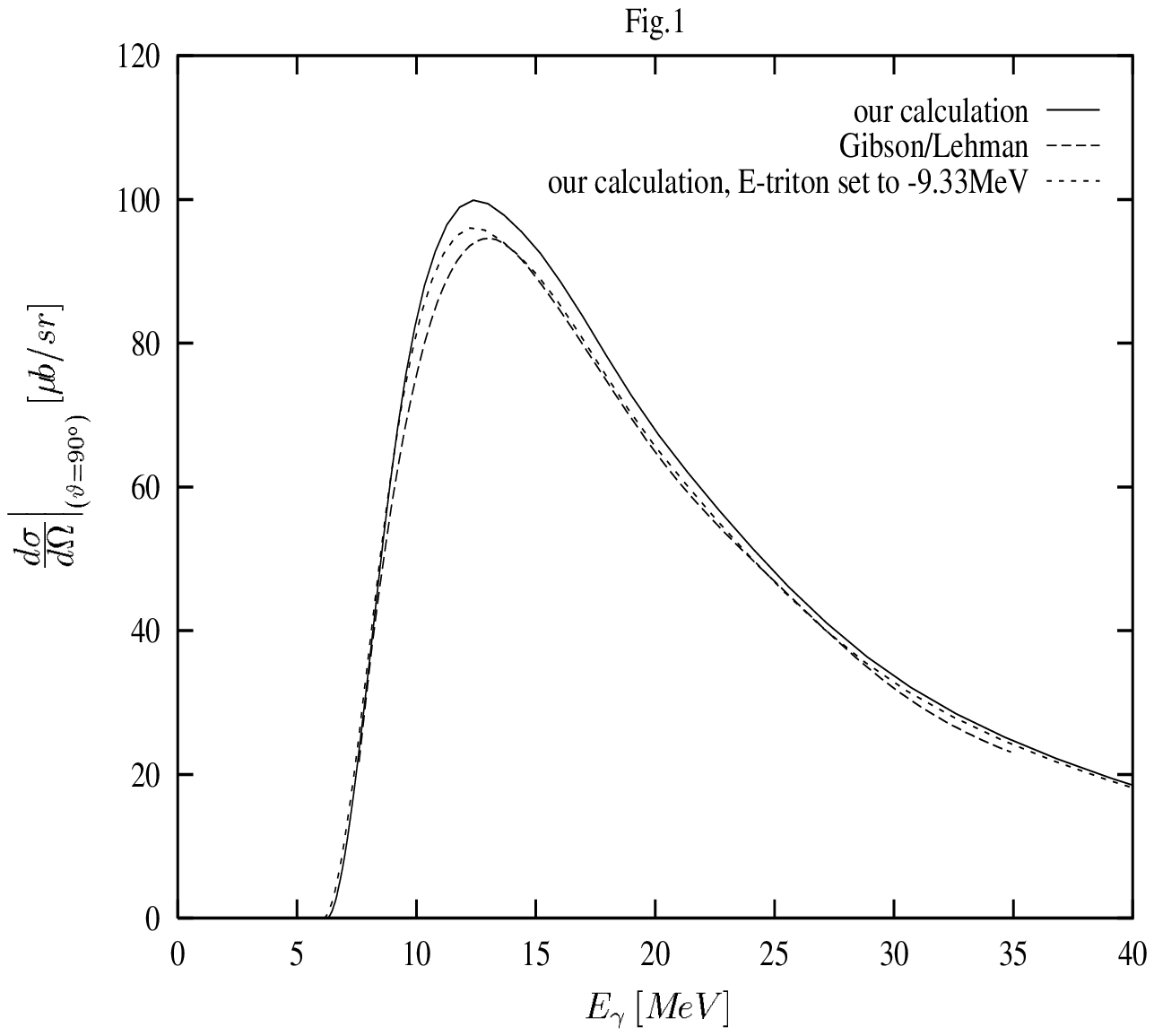,height=10.5cm,width=11.5cm}}
\vspace{1.5cm}
\hbox{\psfig{figure=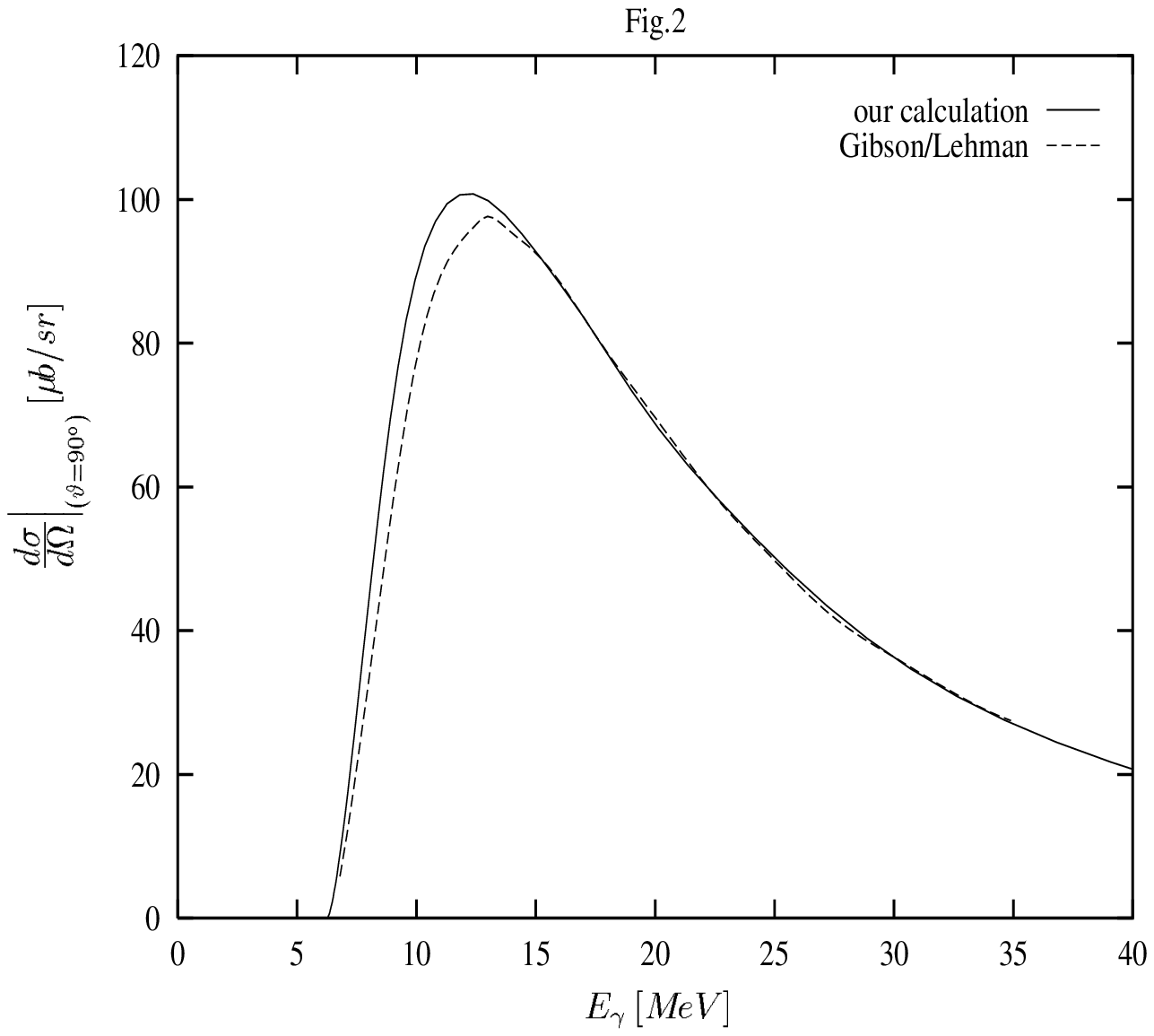,height=10.5cm,width=11.5cm}}
\hbox{\psfig{figure=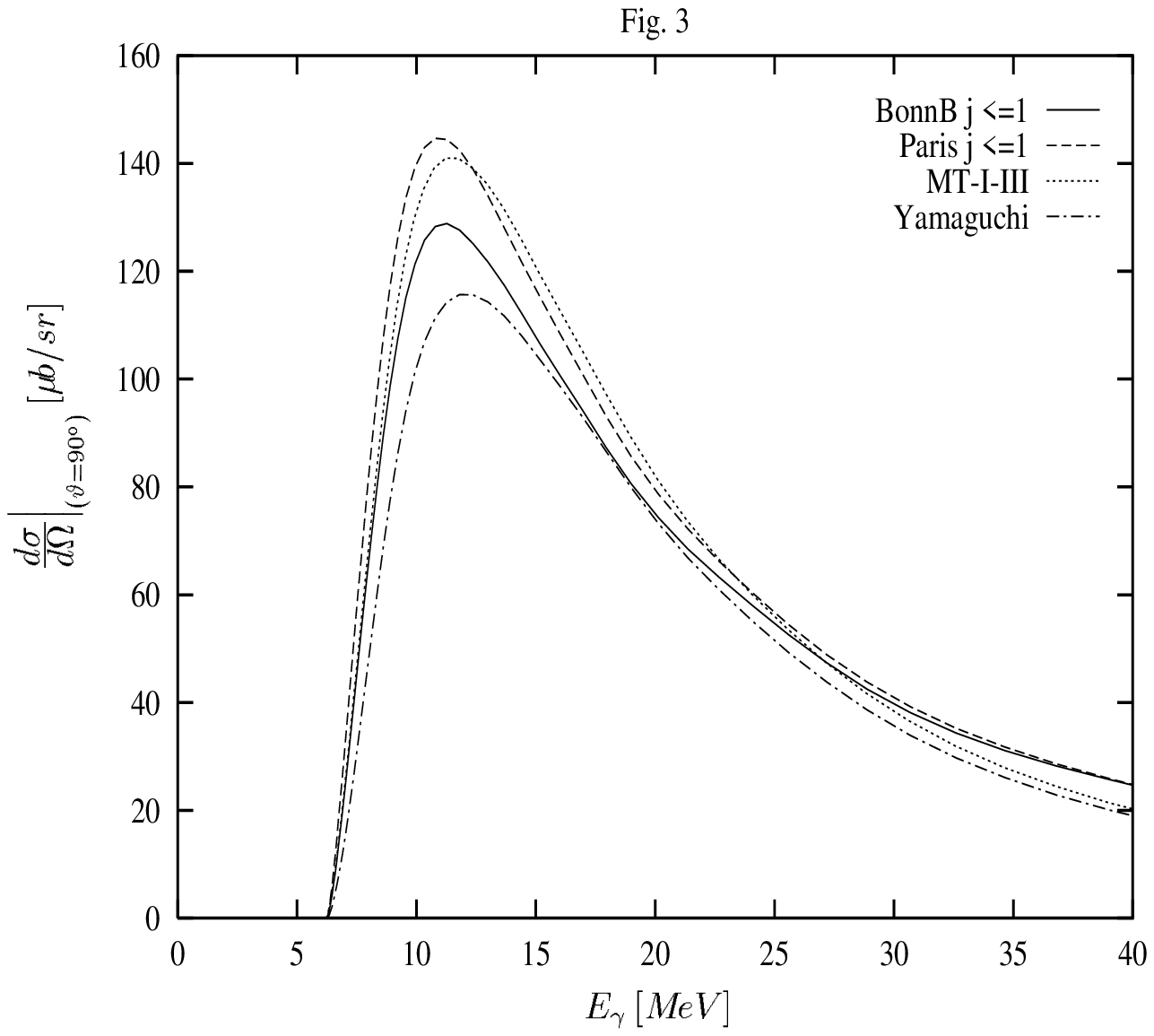,height=10.5cm,width=11.5cm}}
\vspace{1.5cm}
\hbox{\psfig{figure=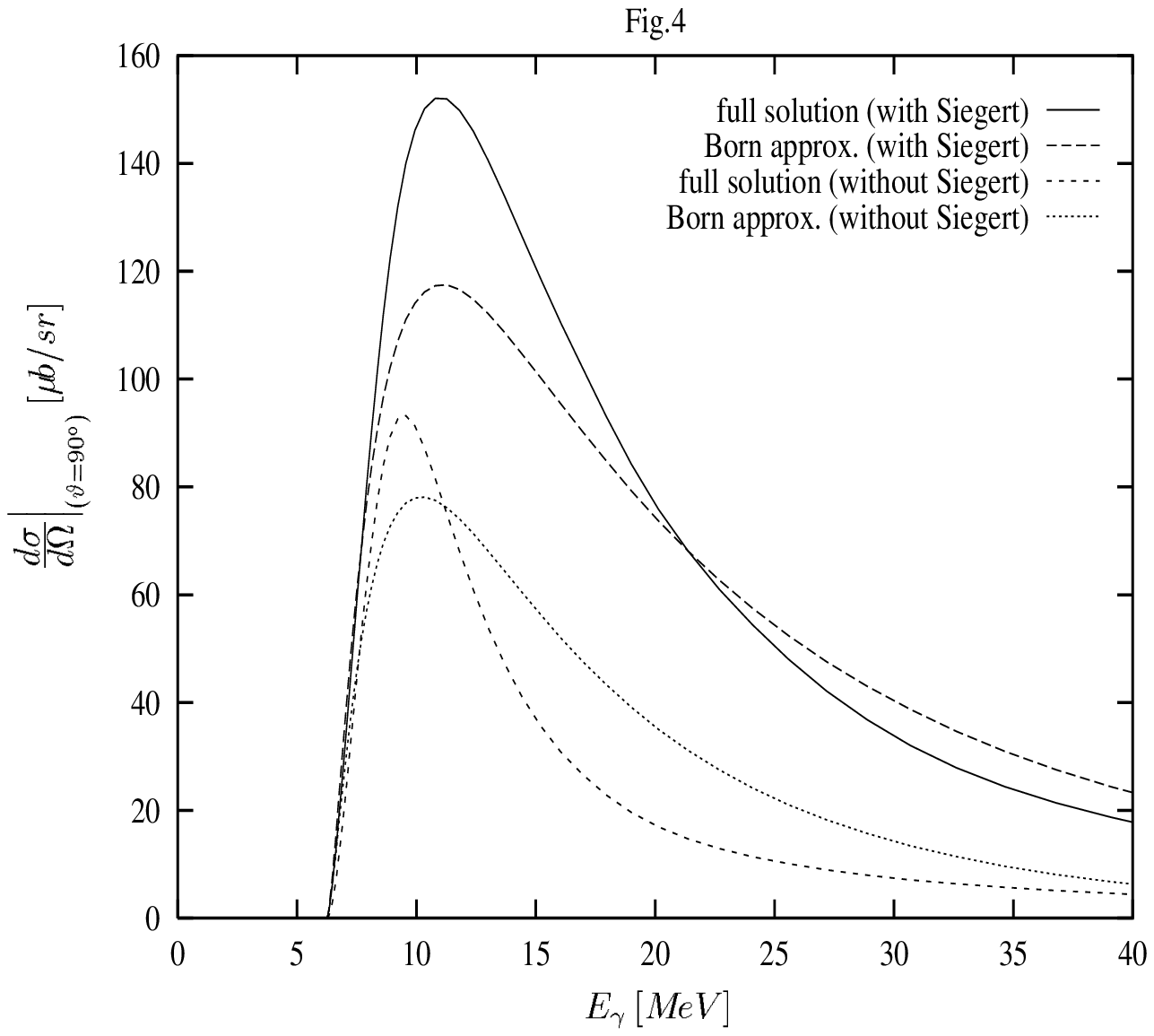,height=10.5cm,width=11.5cm}}
\hbox{\psfig{figure=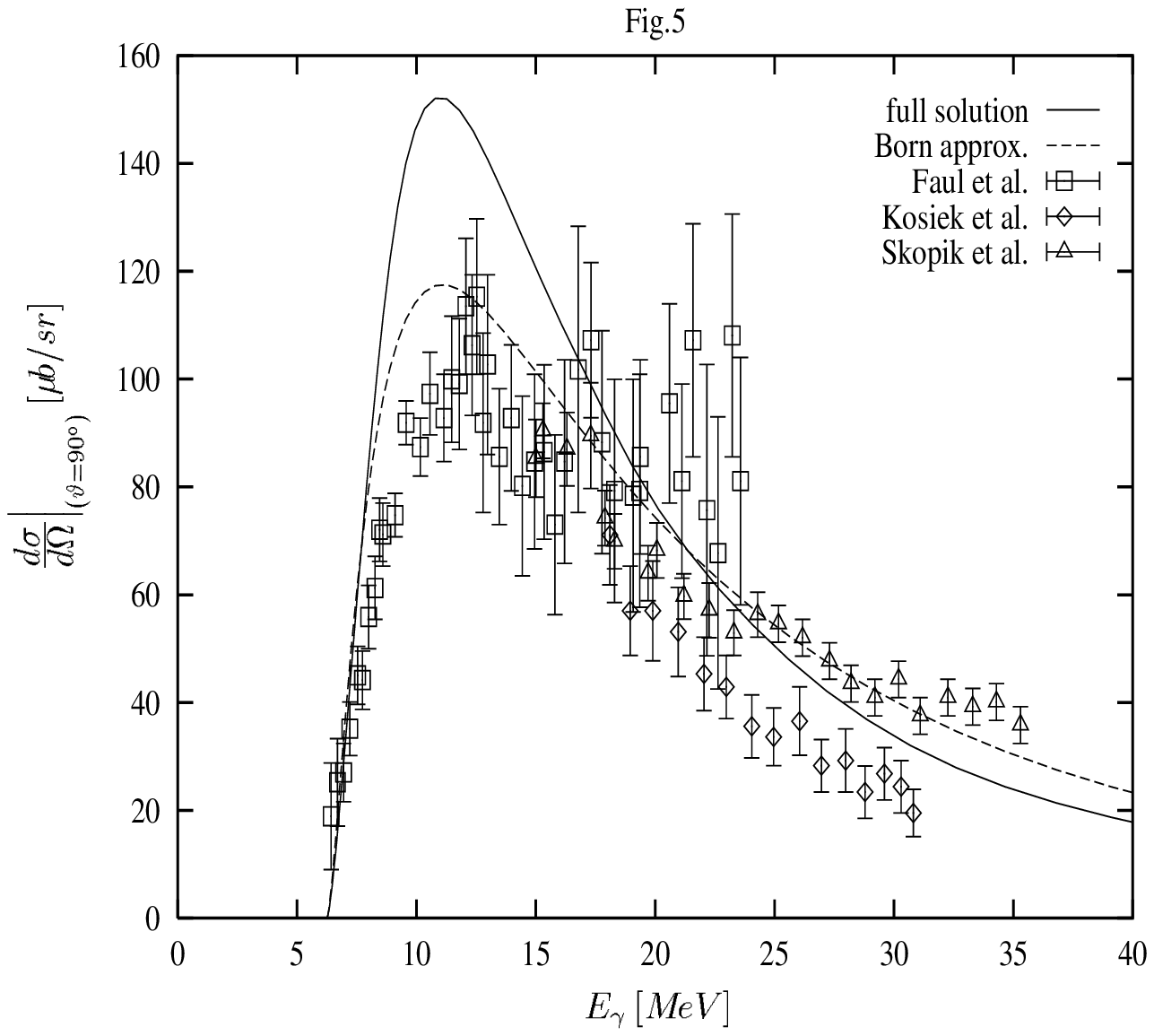,height=10.5cm,width=11.5cm}}
\vspace{1.5cm}
\hbox{\psfig{figure=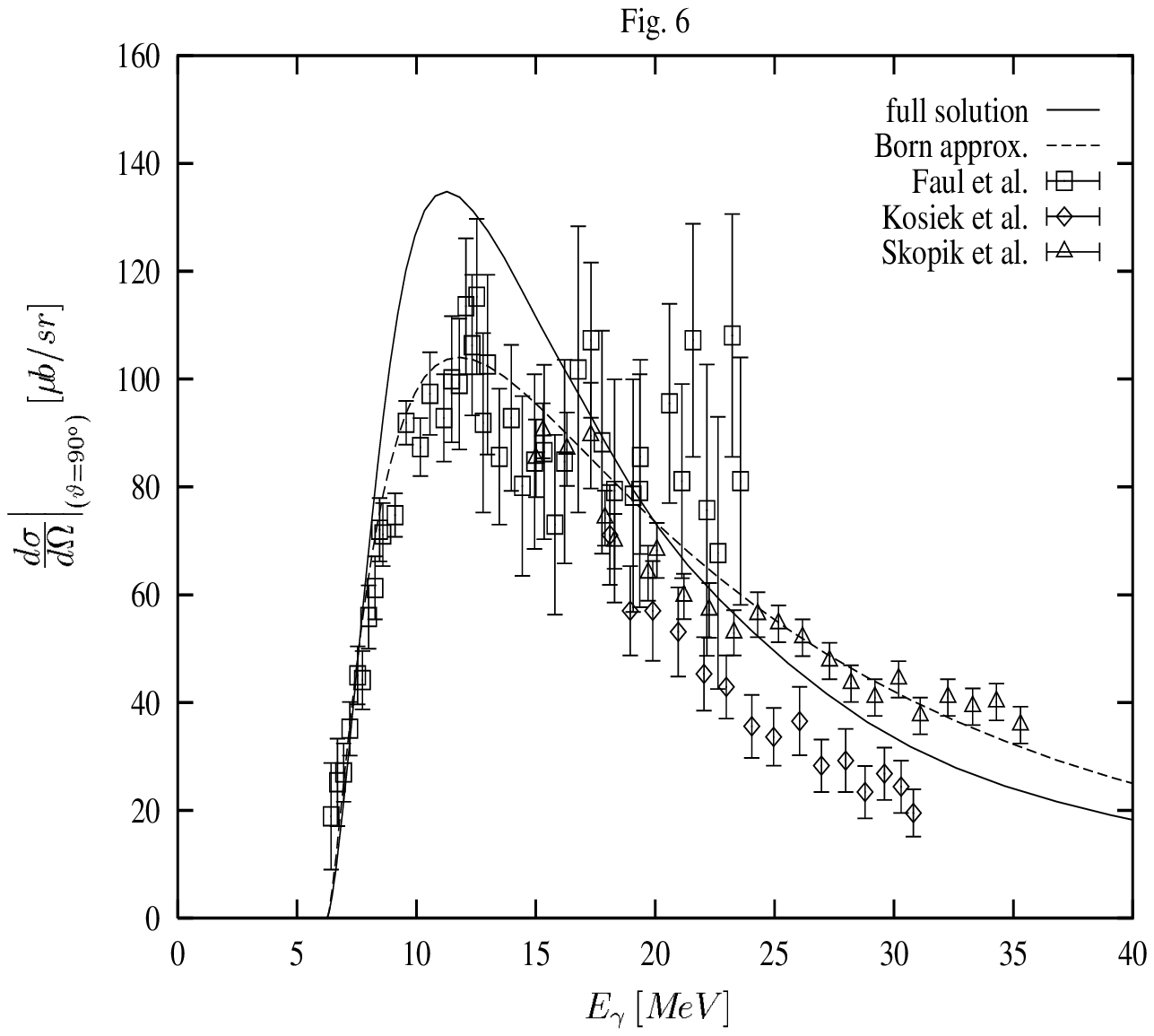,height=10.5cm,width=11.5cm}}
\hbox{\psfig{figure=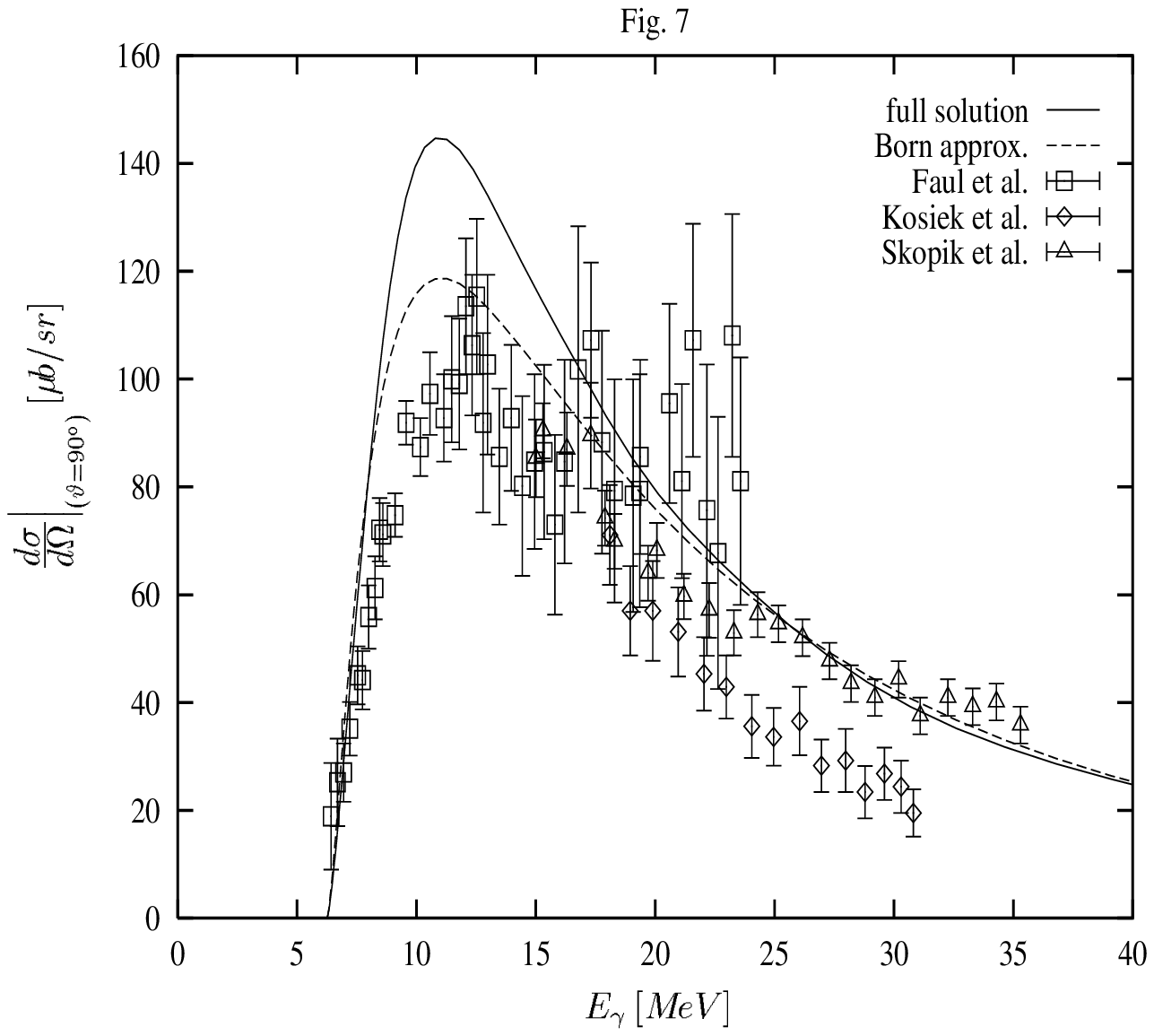,height=10.5cm,width=11.5cm}}
\vspace{1.5cm}
\hbox{\psfig{figure=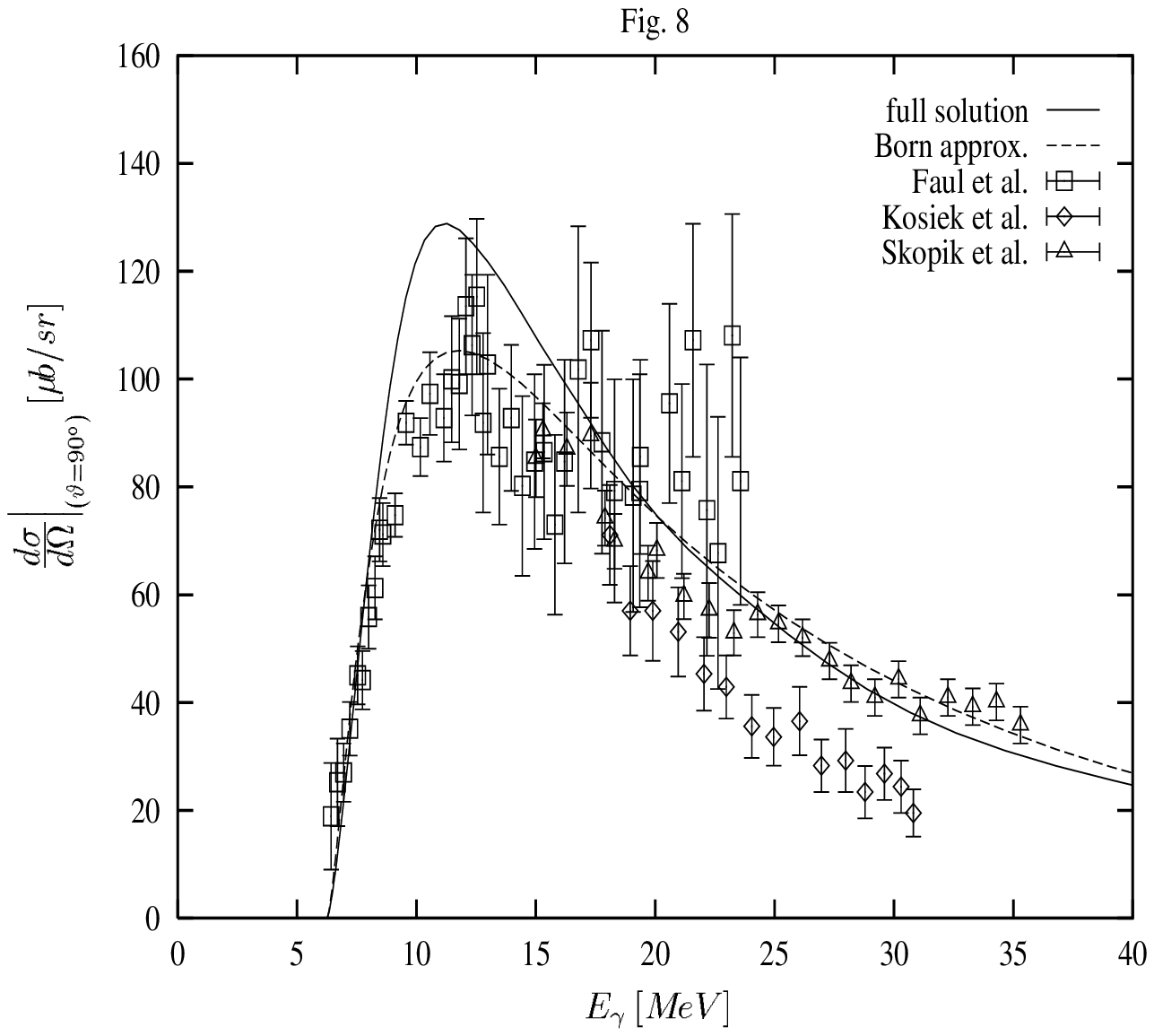,height=10.5cm,width=11.5cm}}

\end{document}